\documentclass [prl,aps,letterpaper,twocolumn,superscriptaddress,amsmath,amssymb,floatfix] {revtex4}

\usepackage{graphicx}
\usepackage{textcomp}
\usepackage{mathptmx} 

\begin{document}

\title{Measurements of Quasiparticle Tunneling Dynamics in a Bandgap-Engineered Transmon Qubit}
\author{L.~Sun}
\affiliation{Department of Physics and Applied Physics, Yale University, New Haven, Connecticut 06520, USA}
\author{L.~DiCarlo}
\affiliation{Department of Physics and Applied Physics, Yale University, New Haven, Connecticut 06520, USA}
\affiliation{Kavli Institute of Nanoscience, Delft University of Technology, Delft, The Netherlands}
\author{M.~D.~Reed}
\author{G.~Catelani}
\affiliation{Department of Physics and Applied Physics, Yale University, New Haven, Connecticut 06520, USA}
\author{Lev~S.~Bishop}
\affiliation{Department of Physics and Applied Physics, Yale University, New Haven, Connecticut 06520, USA}
\affiliation{Joint Quantum Institute and Condensed Matter Theory Center, Department of Physics, University of Maryland, College Park, Maryland 20742, USA}
\author{D.~I.~Schuster}
\affiliation{Department of Physics and Applied Physics, Yale University, New Haven, Connecticut 06520, USA}
\affiliation{Department of Physics and James Franck Institute, University of Chicago, Chicago, Illinois 60637, USA}
\author{B.~R.~Johnson}
\affiliation{Department of Physics and Applied Physics, Yale University, New Haven, Connecticut 06520, USA}
\affiliation{Raytheon BBN Technologies, Cambridge, MA 02138, USA}
\author{Ge~A.~Yang}
\affiliation{Department of Physics and Applied Physics, Yale University, New Haven, Connecticut 06520, USA}
\affiliation{Department of Physics and James Franck Institute, University of Chicago, Chicago, Illinois 60637, USA}
\author{L.~Frunzio}
\author{L.~Glazman}
\author{M.~H.~Devoret}
\author{R.~J.~Schoelkopf}
\affiliation{Department of Physics and Applied Physics, Yale University, New Haven, Connecticut 06520, USA}

\begin{abstract}
We have engineered the bandgap profile of transmon qubits by combining oxygen-doped Al for tunnel junction electrodes and clean Al as quasiparticle traps to investigate energy relaxation due to quasiparticle tunneling. The relaxation time $T_1$ of the qubits is shown to be insensitive to this bandgap engineering. Operating at relatively low $E_J$/$E_C$ makes the transmon transition frequency distinctly dependent on the charge parity, allowing us to detect the quasiparticles tunneling across the qubit junction. Quasiparticle kinetics have been studied by monitoring the frequency switching due to even/odd parity change in real time. It shows the switching time is faster than 10~$\mu$s, indicating quasiparticle-induced relaxation has to be reduced to achieve $T_1$ much longer than 100~$\mu$s.
\end{abstract}

 \maketitle

Quantum information processing based on superconducting qubits has made tremendous progress towards realizing a practical quantum computer in the last few years \cite{DiCarlo1,DiCarlo2,Neeley}. However, the coherence times of superconducting qubits still need to improve to reach the fault-tolerant threshold. For example, the single-qubit gate error rate is limited by qubit decoherence \cite{Lucero,Chow,Paik}. Understanding decoherence mechanisms, in particular those responsible for qubit relaxation, is therefore crucial. Quasiparticles have received significant attention recently as one such possible limiting factor \cite{Martinis,Catelani,Lenander}. At low temperatures, thermal-equilibrium quasiparticles should be irrelevant because their density is exponentially suppressed. In practice, however, non-equilibrium quasiparticles are present from unknown sources \cite{Aumentado,Schreier}. The key question is then whether these non-equilibrium quasiparticles are currently limiting qubit relaxation, and if not, what limit they will ultimately impose. The answer will be relevant to all superconducting qubits.

\begin{figure}
\includegraphics{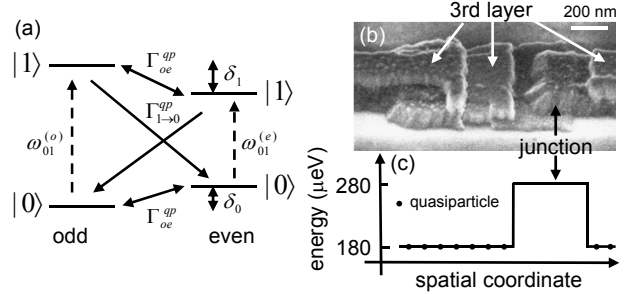}
\caption {(a) Low-energy level structure of a transmon qubit. $\delta_0$ and $\delta_1$: energy difference between odd/even parities for the ground and first excited state respectively, $\delta_1\gg\delta_0$. $\omega_{01}^{(o)}-\omega_{01}^{(e)}=\omega_{oe}\approx\delta_1$ is the charge dispersion. $\Gamma_{oe}^{qp}$ is the odd/even parity switching rate. $\Gamma_{1o\rightarrow0e}^{qp}\approx\Gamma_{1e\rightarrow0o}^{qp}=\Gamma_{1\rightarrow0}^{qp}$ is the qubit relaxation rate because $\omega_{01}^{(o)}\approx\omega_{01}^{(e)}=\omega_{01}\gg\delta_1$. (b) SEM image of a typical device fabricated by three-angle evaporation to engineer the bandgap profile. (c) Expected bandgap profile near the tunnel junction.}
\label{fig:Fig1}
\end{figure}

Quasiparticles in qubit electrodes do not themselves cause significant qubit decoherence. Instead, it is the dissipative and incoherent tunneling of quasiparticles across a Josephson junction that leads to decoherence. The tunneling process is characterized by the quasiparticle current spectral density $S_{qp}(\omega)$. The low frequency component in $S_{qp}(\omega)$ can cause dephasing, but this dephasing channel can be eliminated by reducing the qubit's sensitivity to charge noise, as in flux \cite{Friedman,van}, phase \cite{Martinis2}, and transmon \cite{Koch} qubits. The high frequency component will cause energy relaxation in all types of superconducting qubits. In a transmon qubit, which has the energy level structure shown in Fig.~\ref{fig:Fig1}(a), quasiparticle tunneling leads to two observable rates: the qubit relaxation rate $\Gamma_{1\rightarrow0}^{qp}=|M_{01}|^2S_{qp}(\omega_{01})$ and the odd/even parity switching rate $\Gamma_{oe}^{qp}=|M_{oe}|^2S_{qp}(\omega_{oe})$, where $\omega_{01}/2\pi$ and $\omega_{oe}/2\pi$ are the qubit ground-to-excited and odd-to-even transition frequencies respectively, and $M_{01}$ and $M_{oe}$ are the matrix elements describing the interaction between the qubit and quasiparticles \cite{Catelani2}.

Several approaches may be taken to reduce the detrimental effects of quasiparticles on qubits. One obvious way is to reduce the density of quasiparticles by lowering the temperature, increasing the superconducting bandgap, or suppressing the generation of quasiparticles. Even if the quasiparticles cannot be eliminated, $S_{qp}$ could still be reduced by preventing tunneling of quasiparticles. For example, the bandgap profile could be engineered so that quasiparticle tunneling becomes energetically suppressed. This bandgap engineering approach has been successfully applied to reduce $\Gamma_{oe}^{qp}$ in single-Cooper-pair transistors (SCPTs) \cite{Joyez,Aumentado,Naaman,Court} and superconducting charge qubits \cite{Shaw}.

In this paper, we adopt bandgap engineering to study the relaxation of transmon qubits due to quasiparticles. We measure the qubit relaxation time $T_1$ as a function of temperature of transmon qubits which have significantly different bandgaps. At low temperatures, the saturation of $T_1$ of both transmons at approximately the same level suggests a mechanism other than thermal quasiparticles for qubit relaxation. To investigate if non-equilibrium quasiparticles are limiting $T_1$, we then study the quasiparticle kinetics in a bandgap-engineered transmon operated in the low $E_J$/$E_C$ regime, where $E_J$ and $E_C$ are the Josephson and the charging energy, respectively. Qubit spectroscopy shows two qubit transition frequencies associated with the even- and odd-charge states, demonstrating the presence and tunneling of quasiparticles across the qubit junction. This indicates that with our design, non-equilibrium quasiparticles have not been removed by bandgap engineering, contrary to the results of experiments in SCPTs \cite{Aumentado,Naaman,Court}. We study the parity switching rate $\Gamma_{oe}^{qp}$ in the time domain finding $1/\Gamma_{oe}^{qp}<10~\mu$s. For typical device parameters, the expected ratio of parity switching rate to quasiparticle-induced qubit relaxation rate, $\Gamma_{oe}^{qp}$/$\Gamma_{1\rightarrow 0}^{qp}\sim10-100$ \cite{Catelani2}. Therefore, while we cannot establish quasiparticle tunneling as the dominant source of energy relaxation in our devices, our bound indicates that reducing the quasiparticle-induced decay rate will be necessary to achieve $T_1$ much longer than 100~$\mu$s.

Our transmon qubits are measured using a coplanar waveguide cavity in a conventional circuit quantum electrodynamics architecture \cite{Wallraff}. All devices are measured in a cryogen-free dilution refrigerator with a 20~mK base temperature. Oxygen-doped Al, which has an energy gap $\Delta\approx280~\mu$eV ($T_c\approx1.9$~K) about 60\% higher than clean Al ($\Delta\approx180~\mu$eV, $T_c\approx1.2$~K), is used as the electrodes of Josephson tunnel junctions deposited by standard double-angle evaporation \cite{Dolan}. The oxygen dopants are introduced with a continuous O$_2$ flow during the Al deposition. The same technique has also been used in SCPTs to realize a large bandgap \cite{Aumentado,Naaman}. To create quasiparticle traps, a third layer of clean Al is deposited to cover the whole oxygen-doped Al layers to within $\sim100$~nm from the junctions [Fig.~\ref{fig:Fig1}(b)]. Figure~\ref{fig:Fig1}(c) shows the expected bandgap profile near the junction. This profile is expected to trap quasiparticles away from the tunnel junction. The energy gaps of oxygen-doped layers are determined by independent $T_c$ measurements (not shown) of thin films evaporated under the same nominal conditions as for the tunnel junction.

\begin{figure}
\includegraphics{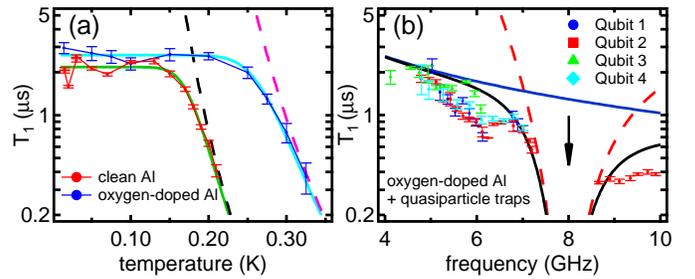}
\caption {(a) Measured qubit relaxation time $T_1$ as a function of temperature for two transmon qubits: one qubit is fabricated with clean Al only with a transition frequency $f_{01}=4.25$~GHz and the other with oxygen-doped Al but without the third layer of quasiparticle traps with $f_{01}=5.16$~GHz. Dashed black and magenta lines are theory for relaxation due to thermal quasiparticles, assuming $\Delta\approx180~\mu$eV and 280~$\mu$eV, respectively \cite{Catelani}. The solid green and cyan lines represent the sum of the theoretical expectations for thermal-equilibrium quasiparticles and the best-fit saturated $T_1$. (b) $T_1$ vs frequency on qubits fabricated with oxygen-doped Al and quasiparticle traps. Dashed red line: Purcell-induced relaxation time; solid blue line: $Q=65,000$; solid black line: a combination of the Purcell effect and a constant $Q$. Neither oxygen-doped Al nor quasiparticle traps improve $T_1$ qualitatively. The black arrow indicates the cavity frequency.}
\label{fig:Fig2}
\end{figure}

However, our bandgap engineering does not appear to affect $T_1$ (see Table~I in Supplemental Material for list of devices). A comparison of $T_1$ as a function of temperature for two representative devices is shown in Fig.~\ref{fig:Fig2}(a): one qubit is fabricated with clean Al only and the other with oxygen-doped Al without the third layer quasiparticle trap. The decrease of $T_1$ with temperatures above 150~mK for the red and 250~mK for the blue curve, respectively, indicates the effect of thermally generated quasiparticles \cite{Paik,Catelani2}. The higher corner temperature confirms that oxygen-doped Al indeed has a larger energy gap. The saturation of $T_1$ at low temperatures for both clean and oxygen-doped Al devices at the same value indicates that thermal quasiparticles are not limiting $T_1$. Figure \ref{fig:Fig2}(b) shows the saturated $T_1$ as a function of frequency for four qubits, each of which has its own flux bias line allowing individual tuning of the qubit frequency \cite{DiCarlo2}, fabricated with both oxygen-doped Al and quasiparticle traps described earlier. The qubit lifetimes are limited to a quality factor $Q\sim65,000$, similar to previously reported $T_1$ on qubits fabricated with clean Al only \cite{Houck}.

\begin{figure}
\includegraphics{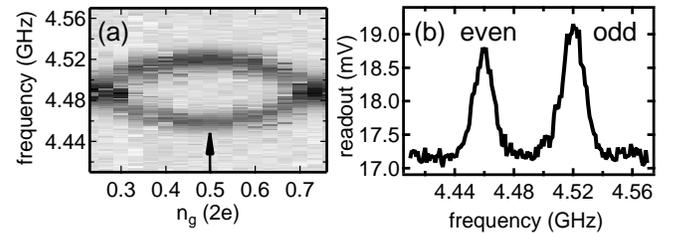}
\caption {(a) Spectroscopy of a qubit with engineered bandgap as a function of gate-induced charge $n_g$. The gate voltage is applied to the qubit through a bias tee at the cavity input port. For each pixel, a gaussian pulse ($\sigma=20$~ns, corresponding to a $\pi$ pulse on resonance) is applied at the indicated frequency and the qubit is immediately measured. Each pixel is average of 5000 repetitions (50~ms). Darker pixels correspond to higher homodyne readout voltages which are proportional to the probability of the qubit in the excited state. An ``eye"-shaped pattern indicates charge-$e$ jumps associated with the tunneling of non-equilibrium quasiparticles. (b) Cross-section of (a) at $n_g=0.5$ [black arrow in (a)]. The charge dispersion is 60~MHz. We refer to the lower (upper) frequency branch as the even (odd) parity peak.}
\label{fig:Fig3}
\end{figure}

Non-equilibrium quasiparticles have been observed in the bandgap-engineered devices as will be shown later, so here we will only focus on the effect on $T_1$ from those non-equilibrium quasiparticles. The lack of qualitative improvements of $T_1$ suggests two possible scenarios: either $T_1$ is not limited by non-equilibrium quasiparticles, or the quasiparticle contribution to $T_1$ has not been affected by our bandgap engineering. In either case, we wish to find the quasiparticle-induced qubit relaxation rate $\Gamma_{1\rightarrow0}^{qp}$. However, this rate cannot be measured directly in the presence of other sources of relaxation, so we instead measure the dynamics of quasiparticle tunneling $\Gamma_{oe}^{qp}$ in the time domain, from which we can estimate $\Gamma_{1\rightarrow0}^{qp}$ \cite{Catelani2}. In order to measure $\Gamma_{oe}^{qp}$, we operate the transmon qubit in the low $E_J/E_C$ regime where the qubit spectrum will have two distinct parities caused by quasiparticle tunneling across the junction. Thus, $\Gamma_{oe}^{qp}$ can be studied by monitoring the dynamics of one particular parity in real time.

We have fabricated a single transmon qubit with both oxygen-doped Al and quasiparticle traps, and operate it at $E_J/E_C\approx12.5$ by tuning the qubit frequency to $f_{01}\approx4.5$~GHz, where $T_1=2~\mu$s. To protect the qubit from spontaneous emission \cite{Houck} through the low-$Q$ cavity ($Q=500$), a Purcell filter is also integrated on chip \cite{Reed1}. At this large detuning from the cavity frequency $f_c=8.072$~GHz, a direct readout of the qubit state is difficult because of the weak dispersive interaction between the qubit and the cavity. Instead, to enhance readout fidelity, the qubit is pulsed to $f_{01}\approx7$~GHz prior to measurement. By making use of the high-power Jaynes-Cummings readout \cite{Reed2}, we achieve a single-shot fidelity $F=42\%$.

The presence and tunneling of non-equilibrium quasiparticles is demonstrated by the clearly observed ``eye" pattern in the qubit spectroscopy as a function of gate-induced charge $n_g$ [Fig.~\ref{fig:Fig3}(a)] \cite{Schreier}. The two peaks in Fig.~\ref{fig:Fig3}(b) have almost equal height, implying no preferred parity, and both parity switching times $\tau_o$ (odd-to-even) and $\tau_e$ (even-to-odd) shorter than the 50~ms averaging time at each pixel.

\begin{figure}
\includegraphics{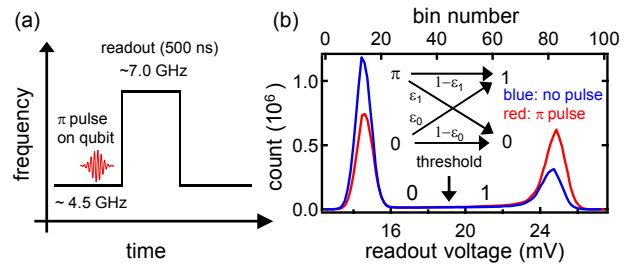}
\caption {(a) Schematic of the measurement. A selective $\pi$ pulse is applied to one of the two parity peaks in Fig.~\ref{fig:Fig3} followed by an immediate readout of the qubit state at about 7 GHz where readout fidelity is improved. Lines: the flux pulse sequence. The process is repeated every 10 $\mu$s. (b) Histogram of the readout at the odd parity peak in Fig.~\ref{fig:Fig3}(b). A threshold $V_{th}=19$~mV is chosen to digitize the readout. Inset: Schematic of an imperfect readout with false positives (negatives) $\epsilon_0$ ($\epsilon_1$).}
\label{fig:Fig4}
\end{figure}

To study the dynamics of quasiparticle tunneling, a selective $\pi$ pulse is applied repeatedly every 10~$\mu$s at one of two parity frequencies (for simplicity, we choose the odd parity frequency for all data presented here), immediately followed by a measurement of the qubit state [see Fig.~\ref{fig:Fig4}(a)]. This selective $\pi$ pulse will excite the qubit only when the parity is odd. We note that due to the smallness of $\delta_0$ [Fig.~\ref{fig:Fig1}(a)], the selective $\pi$ pulse will also cover the transition $\left|0,even\right>$ to $\left|1,odd\right>$, but for non-degenerate quasiparticles the probability of this unwanted transition is small compared to that of the direct photon transition $\left|0,odd\right>$ to $\left|1,odd\right>$. Thus, by interrogating the qubit state after the $\pi$ pulse, the qubit parity can be inferred. To ensure that the qubit begins in the ground state and that the qubit frequency has stabilized after the rapid tuning from 7 to 4.5~GHz at the beginning of each experimental cycle, the repetition rate is set to $t_s=10~\mu$s $= 5 T_1$. Because quasiparticles tunnel across the qubit junction randomly, a random telegraph signal (RTS) is expected. To overcome low single-shot readout fidelity, we perform a Fourier transform of the time domain data to study the power spectral density (PSD) to better extract the RTS information. Background charge motion limits the experiment duration since the qubit frequency shifts noticeably every few minutes. In total, twelve million measurements are recorded and Fourier transformed.

The stability of the readout is particularly important for this measurement, and so to minimize the drift during the experiment, the readout result is digitized by thresholding the measurement results [Fig.~\ref{fig:Fig4}(b)]. From each measurement, one bit of information is extracted. False positives $\epsilon_0$ and negatives $\epsilon_1$ of the readout [Fig.~\ref{fig:Fig4}(b) inset] will reduce the RTS amplitude. Note that due to the switching of the qubit between the two parities, $\epsilon_0$ and $\epsilon_1$ cannot be extracted directly from Fig.~\ref{fig:Fig4}(b), but can easily be obtained by combining histograms of the readout at the even parity peak in Fig.~\ref{fig:Fig3}(b) (see Supplemental Material for details). We find $\epsilon_0\approx\epsilon_1=0.29$, as well as the probability of the qubit at the odd parity $P_{odd}=56\%$. Thus the readout fidelity is $F=1-\epsilon_0-\epsilon_1\approx42\%$.

\begin{figure}[t]
\includegraphics{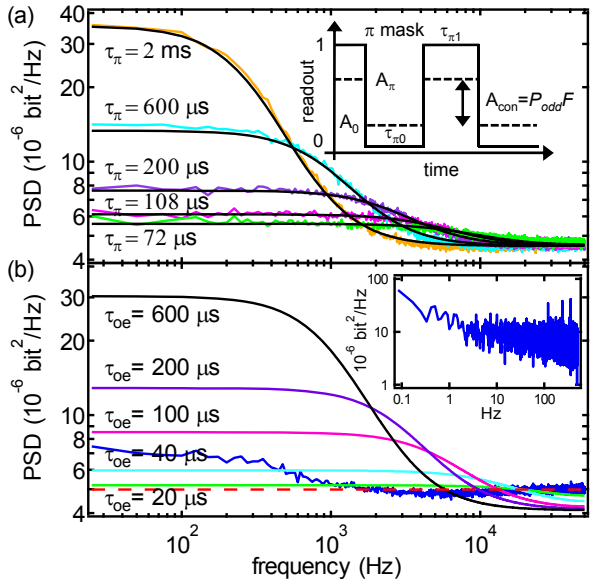}
\caption {(a) Power spectral densities of control experiments using a $\pi$ mask with different time constants $\tau_\pi=\tau_{\pi 0}+\tau_{\pi 1}$. $\tau_{\pi 0}=\tau_{\pi 1}$ are the time constants associated with the $\pi$ mask. The color curves are data and black curves are theoretical predictions. Inset: schematic of the control experiment with a $\pi$ mask, an RTS with specified time constants and generated by an FPGA. The expected RTS amplitude is reduced to $A_{con}=A_\pi-A_0=P_{odd}F$, where $A_\pi$ and $A_0$ are the statistical averages of the readout during the $\pi$ mask assuming parity switches fast compared to $\tau_\pi$, $P_{odd}$ is the probability of the qubit to be in the odd parity state, and $F$ is the readout fidelity. (b) Measured PSDs and theoretical predictions for the direct experiment on $\Gamma_{oe}^{qp}$ without the $\pi$ mask. Blue curve is the measured Fourier spectrum and the color curves are theoretical predictions for different qubit parity switching times $\tau_{oe}=\tau_o+\tau_e$, where $\tau_o$ and $\tau_e$ are the odd-to-even and even-to-odd switching time, respectively. The red dashed line is the white noise spectrum in the limit $\tau_o$, $\tau_e<t_s$, the sampling time. The near absence of any deviation from white noise at high frequencies indicates $\tau_o, \tau_e<10~\mu$s. Inset: The PSD at low frequency does not show a Lorentzian plateau.}
\label{fig:Fig5}
\end{figure}

We first test the sensitivity of the experiment to fluctuations of parity by applying a ``$\pi$ mask'' to the measurement system to imitate the success and failure of a $\pi$ pulse on the qubit. The $\pi$ mask, generated by a field programmable gate array (FPGA), is a pseudo-random control sequence which enables or disables a $\pi$ pulse applied to the qubit immediately prior to readout. It generates an RTS with a specified time constant $\tau_\pi=\tau_{\pi 0}+\tau_{\pi 1}$ and a 50\% duty cycle ($\tau_{\pi 0}=\tau_{\pi 1}$). If the parity switches fast compared to $\tau_{\pi}$ (as will be confirmed later), the RTS amplitude in the readout is reduced to $A_{con}=P_{odd}F$ due to the parity duty cycle and the finite readout fidelity [Fig.~\ref{fig:Fig5}(a) inset, also see Supplementary Material]. Figure~\ref{fig:Fig5}(a) shows the PSDs of the control experiments with a $\pi$ mask with different time constants $\tau_\pi=$2~ms, 600~$\mu$s, 200~$\mu$s, 108~$\mu$s, and 72~$\mu$s, respectively. All measured PSDs agree very well with theory (see Supplemental Material for details).

We now turn to the measurement of the parity switching rate $\Gamma_{oe}^{qp}$ of the qubit, showing in Fig.~\ref{fig:Fig5}(b) the PSD obtained without the $\pi$ mask. The measured spectrum is almost flat and white-noise like, suggesting fast qubit parity switching. An upper bound can be placed on $\Gamma_{oe}^{qp}$ by comparing with the theoretical predictions for the observed duty cycle $P_{odd}=\tau_o/(\tau_o+\tau_e) =56\%$ and different time constants, $\tau_o$ and $\tau_e$. We note that the theoretical predictions in Fig.~\ref{fig:Fig5}(b) assume an RTS amplitude $A=F$, almost twice as large as that in the calibration experiment, because the RTS is just from parity switching (see the Supplemental Material). Even at $\tau_{oe}=\tau_o+\tau_e=20~\mu$s, there is still some deviation of the theoretical prediction from data. We infer from this that qubit parity switching time is faster than our repetition time, $\tau_o\sim\tau_e<10~\mu$s. This upper bound has been lowered by more than three orders of magnitude from the previous estimate in Ref.~\onlinecite{Schreier}. In Fig.~\ref{fig:Fig5}(b), there is a small rise in the spectrum below $\sim1$~kHz, but its amplitude is an order of magnitude smaller than the theoretical prediction at $\tau_{oe}=$2~ms with the same corner frequency, excluding quasiparticle tunneling as the cause. A slow background charge motion can contribute to this low-frequency plateau. An analysis of the spectrum down to 0.1~Hz with the absence of Lorentzian plateau [Fig.~\ref{fig:Fig5}(b) inset] confirms that no slow RTS process is missed.

The upper bound of $\tau_o\sim\tau_e<10~\mu$s places an upper bound on the quasiparticle-induced qubit relaxation time $1/\Gamma_{1\rightarrow 0}^{qp}$. Recent theory \cite{Catelani2} predicts that the quasiparticle-induced qubit decay rate is much slower than the even/odd parity switching rate. For current device parameters, $\Gamma_{oe}^{qp}\approx1/\tau_o$ and $\Gamma_{oe}^{qp}$/$\Gamma_{1\rightarrow 0}^{qp}\sim30$, implying a qubit relaxation time $1/\Gamma_{1\rightarrow 0}^{qp}<300~\mu$s. Combining this ratio with the measured $T_1=2~\mu$s gives a lower bound $\tau_o\sim\tau_e>100$~ns. These results do not allow us to determine whether quasiparticles dominate the energy relaxation of current transmon qubits. However, it is evident that quasiparticle tunneling will need to be reduced in order to extend $T_1$ beyond $\sim100~\mu$s.

We have engineered the bandgap profile of transmon qubits by combining oxygen-doped Al for tunnel junction electrodes and clean Al as quasiparticle traps in an attempt to reduce the qubit relaxation due to quasiparticles. However, the measured qubit relaxation time is found to be insensitive to the size of the superconducting gap and to quasiparticle traps. On the other hand, the nondegradation of the qubit relaxation time suggests no measurable change of the junction quality from oxygen-doped Al. Non-equilibrium quasiparticles leading to the charge-parity switching have been observed in all devices. Moreover, the quasiparticle-induced parity switching is shown to be faster than 10~$\mu$s, an upper bound limited by the detection bandwidth. This fast parity switching rate, different from the results of other experiments in SCPTs, could be due to three things. First, the size of the transmon electrodes is much larger and the quasiparticle density may depend on the electrode size. Second, the qubit readout might generate quasiparticles or stimulate the tunneling of quasiparticles between electrodes. A third possibility is that the interface between the oxygen-doped and clean Al layers might not be transparent enough. Although our results neither prove nor disprove that non-equilibrium quasiparticles limit $T_1$ in transmon qubits, it does indicate that quasiparticle-induced energy relaxation must be reduced in the future to achieve $T_1$ much longer than 100~$\mu$s.

We thank E.~Ginossar, H.~Paik, A.~Sears, G.~Kirchmair, S.~Shankar, M.~Hatridge, and S.~Girvin for valuable discussions. L.F. acknowledges partial support from CNR-Istituto di Cibernetica. L.D.C. acknowledges partial support from NWO. This research was funded by the Office of the Director of National Intelligence (ODNI), Intelligence Advanced Research Projects Activity (IARPA), through the Army Research Office. All statements of fact, opinion or conclusions contained herein are those of the authors and should not be construed as representing the official views or policies of IARPA, the ODNI, or the U.S. Government.


\begin{thebibliography}{25}
\expandafter\ifx\csname natexlab\endcsname\relax\def\natexlab#1{#1}\fi
\expandafter\ifx\csname bibnamefont\endcsname\relax
  \def\bibnamefont#1{#1}\fi
\expandafter\ifx\csname bibfnamefont\endcsname\relax
  \def\bibfnamefont#1{#1}\fi
\expandafter\ifx\csname citenamefont\endcsname\relax
  \def\citenamefont#1{#1}\fi
\expandafter\ifx\csname url\endcsname\relax
  \def\url#1{\texttt{#1}}\fi
\expandafter\ifx\csname urlprefix\endcsname\relax\def\urlprefix{URL }\fi
\providecommand{\bibinfo}[2]{#2}
\providecommand{\eprint}[2][]{\url{#2}}

\bibitem[{\citenamefont{DiCarlo et~al.}(2009)\citenamefont{DiCarlo, Chow,
  Gambetta, Bishop, Johnson, Schuster, Majer, Blais, Frunzio, Girvin
  et~al.}}]{DiCarlo1}
\bibinfo{author}{\bibfnamefont{L.}~\bibnamefont{DiCarlo}},
  \bibinfo{author}{\bibfnamefont{J.~M.} \bibnamefont{Chow}},
  \bibinfo{author}{\bibfnamefont{J.~M.} \bibnamefont{Gambetta}},
  \bibinfo{author}{\bibfnamefont{L.~S.} \bibnamefont{Bishop}},
  \bibinfo{author}{\bibfnamefont{B.~R.} \bibnamefont{Johnson}},
  \bibinfo{author}{\bibfnamefont{D.~I.} \bibnamefont{Schuster}},
  \bibinfo{author}{\bibfnamefont{J.}~\bibnamefont{Majer}},
  \bibinfo{author}{\bibfnamefont{A.}~\bibnamefont{Blais}},
  \bibinfo{author}{\bibfnamefont{L.}~\bibnamefont{Frunzio}},
  \bibinfo{author}{\bibfnamefont{S.~M.} \bibnamefont{Girvin}},
  \bibnamefont{et~al.}, \bibinfo{journal}{Nature}
  \textbf{\bibinfo{volume}{460}}, \bibinfo{pages}{240} (\bibinfo{year}{2009}).

\bibitem[{\citenamefont{DiCarlo et~al.}(2010)\citenamefont{DiCarlo, Reed, Sun,
  Johnson, Chow, Gambetta, Frunzio, Girvin, Devoret, and
  Schoelkopf}}]{DiCarlo2}
\bibinfo{author}{\bibfnamefont{L.}~\bibnamefont{DiCarlo}},
  \bibinfo{author}{\bibfnamefont{M.~D.} \bibnamefont{Reed}},
  \bibinfo{author}{\bibfnamefont{L.}~\bibnamefont{Sun}},
  \bibinfo{author}{\bibfnamefont{B.~R.} \bibnamefont{Johnson}},
  \bibinfo{author}{\bibfnamefont{J.~M.} \bibnamefont{Chow}},
  \bibinfo{author}{\bibfnamefont{J.~M.} \bibnamefont{Gambetta}},
  \bibinfo{author}{\bibfnamefont{L.}~\bibnamefont{Frunzio}},
  \bibinfo{author}{\bibfnamefont{S.~M.} \bibnamefont{Girvin}},
  \bibinfo{author}{\bibfnamefont{M.~H.} \bibnamefont{Devoret}},
  \bibnamefont{and} \bibinfo{author}{\bibfnamefont{R.~J.}
  \bibnamefont{Schoelkopf}}, \bibinfo{journal}{Nature}
  \textbf{\bibinfo{volume}{467}}, \bibinfo{pages}{574} (\bibinfo{year}{2010}).

\bibitem[{\citenamefont{Neeley et~al.}(2010)\citenamefont{Neeley, Bialczak,
  Lenander, Lucero, Mariantoni, O'Connell, Sank, Wang, Weides, Wenner
  et~al.}}]{Neeley}
\bibinfo{author}{\bibfnamefont{M.}~\bibnamefont{Neeley}},
  \bibinfo{author}{\bibfnamefont{R.~C.} \bibnamefont{Bialczak}},
  \bibinfo{author}{\bibfnamefont{M.}~\bibnamefont{Lenander}},
  \bibinfo{author}{\bibfnamefont{E.}~\bibnamefont{Lucero}},
  \bibinfo{author}{\bibfnamefont{M.}~\bibnamefont{Mariantoni}},
  \bibinfo{author}{\bibfnamefont{A.~D.} \bibnamefont{O'Connell}},
  \bibinfo{author}{\bibfnamefont{D.}~\bibnamefont{Sank}},
  \bibinfo{author}{\bibfnamefont{H.}~\bibnamefont{Wang}},
  \bibinfo{author}{\bibfnamefont{M.}~\bibnamefont{Weides}},
  \bibinfo{author}{\bibfnamefont{J.}~\bibnamefont{Wenner}},
  \bibnamefont{et~al.}, \bibinfo{journal}{Nature}
  \textbf{\bibinfo{volume}{467}}, \bibinfo{pages}{570} (\bibinfo{year}{2010}).

\bibitem[{\citenamefont{Lucero et~al.}(2008)\citenamefont{Lucero, Hofheinz,
  Ansmann, Bialczak, Katz, Neeley, O'Connell, Wang, Cleland, and
  Martinis}}]{Lucero}
\bibinfo{author}{\bibfnamefont{E.}~\bibnamefont{Lucero}},
  \bibinfo{author}{\bibfnamefont{M.}~\bibnamefont{Hofheinz}},
  \bibinfo{author}{\bibfnamefont{M.}~\bibnamefont{Ansmann}},
  \bibinfo{author}{\bibfnamefont{R.~C.} \bibnamefont{Bialczak}},
  \bibinfo{author}{\bibfnamefont{N.}~\bibnamefont{Katz}},
  \bibinfo{author}{\bibfnamefont{M.}~\bibnamefont{Neeley}},
  \bibinfo{author}{\bibfnamefont{A.~D.} \bibnamefont{O'Connell}},
  \bibinfo{author}{\bibfnamefont{H.}~\bibnamefont{Wang}},
  \bibinfo{author}{\bibfnamefont{A.~N.} \bibnamefont{Cleland}},
  \bibnamefont{and} \bibinfo{author}{\bibfnamefont{J.~M.}
  \bibnamefont{Martinis}}, \bibinfo{journal}{Phys. Rev. Lett.}
  \textbf{\bibinfo{volume}{100}}, \bibinfo{pages}{247001}
  (\bibinfo{year}{2008}).

\bibitem[{\citenamefont{Chow et~al.}(2009)\citenamefont{Chow, Gambetta,
  Tornberg, Koch, Bishop, Houck, Johnson, Frunzio, Girvin, and
  Schoelkopf}}]{Chow}
\bibinfo{author}{\bibfnamefont{J.~M.} \bibnamefont{Chow}},
  \bibinfo{author}{\bibfnamefont{J.~M.} \bibnamefont{Gambetta}},
  \bibinfo{author}{\bibfnamefont{L.}~\bibnamefont{Tornberg}},
  \bibinfo{author}{\bibfnamefont{J.}~\bibnamefont{Koch}},
  \bibinfo{author}{\bibfnamefont{L.~S.} \bibnamefont{Bishop}},
  \bibinfo{author}{\bibfnamefont{A.~A.} \bibnamefont{Houck}},
  \bibinfo{author}{\bibfnamefont{B.~R.} \bibnamefont{Johnson}},
  \bibinfo{author}{\bibfnamefont{L.}~\bibnamefont{Frunzio}},
  \bibinfo{author}{\bibfnamefont{S.~M.} \bibnamefont{Girvin}},
  \bibnamefont{and} \bibinfo{author}{\bibfnamefont{R.~J.}
  \bibnamefont{Schoelkopf}}, \bibinfo{journal}{Phys. Rev. Lett.}
  \textbf{\bibinfo{volume}{102}}, \bibinfo{pages}{090502}
  (\bibinfo{year}{2009}).

\bibitem[{\citenamefont{Paik et~al.}(2011)\citenamefont{Paik, Schuster, Bishop,
  Kirchmair, Catelani, Sears, Johnson, Reagor, Frunzio, Glazman et~al.}}]{Paik}
\bibinfo{author}{\bibfnamefont{H.}~\bibnamefont{Paik}},
  \bibinfo{author}{\bibfnamefont{D.~I.} \bibnamefont{Schuster}},
  \bibinfo{author}{\bibfnamefont{L.~S.} \bibnamefont{Bishop}},
  \bibinfo{author}{\bibfnamefont{G.}~\bibnamefont{Kirchmair}},
  \bibinfo{author}{\bibfnamefont{G.}~\bibnamefont{Catelani}},
  \bibinfo{author}{\bibfnamefont{A.~P.} \bibnamefont{Sears}},
  \bibinfo{author}{\bibfnamefont{B.~R.} \bibnamefont{Johnson}},
  \bibinfo{author}{\bibfnamefont{M.~J.} \bibnamefont{Reagor}},
  \bibinfo{author}{\bibfnamefont{L.}~\bibnamefont{Frunzio}},
  \bibinfo{author}{\bibfnamefont{L.}~\bibnamefont{Glazman}},
  \bibnamefont{et~al.}, \bibinfo{journal}{Phys. Rev. Lett.}
  \textbf{\bibinfo{volume}{107}}, \bibinfo{pages}{240501}
  (\bibinfo{year}{2011}).

\bibitem[{\citenamefont{Martinis et~al.}(2009)\citenamefont{Martinis, Ansmann,
  and Aumentado}}]{Martinis}
\bibinfo{author}{\bibfnamefont{J.~M.} \bibnamefont{Martinis}},
  \bibinfo{author}{\bibfnamefont{M.}~\bibnamefont{Ansmann}}, \bibnamefont{and}
  \bibinfo{author}{\bibfnamefont{J.}~\bibnamefont{Aumentado}},
  \bibinfo{journal}{Phys. Rev. Lett.} \textbf{\bibinfo{volume}{103}},
  \bibinfo{pages}{097002} (\bibinfo{year}{2009}).

\bibitem[{\citenamefont{Catelani
  et~al.}(2011{\natexlab{a}})\citenamefont{Catelani, Koch, Frunzio, Schoelkopf,
  Devoret, and Glazman}}]{Catelani}
\bibinfo{author}{\bibfnamefont{G.}~\bibnamefont{Catelani}},
  \bibinfo{author}{\bibfnamefont{J.}~\bibnamefont{Koch}},
  \bibinfo{author}{\bibfnamefont{L.}~\bibnamefont{Frunzio}},
  \bibinfo{author}{\bibfnamefont{R.~J.} \bibnamefont{Schoelkopf}},
  \bibinfo{author}{\bibfnamefont{M.~H.} \bibnamefont{Devoret}},
  \bibnamefont{and} \bibinfo{author}{\bibfnamefont{L.~I.}
  \bibnamefont{Glazman}}, \bibinfo{journal}{Phys. Rev. Lett.}
  \textbf{\bibinfo{volume}{106}}, \bibinfo{pages}{077002}
  (\bibinfo{year}{2011}{\natexlab{a}}).

\bibitem[{\citenamefont{Lenander et~al.}(2011)\citenamefont{Lenander, Wang,
  Bialczak, Lucero, Mariantoni, Neeley, O'Connell, Sank, Weides, Wenner
  et~al.}}]{Lenander}
\bibinfo{author}{\bibfnamefont{M.}~\bibnamefont{Lenander}},
  \bibinfo{author}{\bibfnamefont{H.}~\bibnamefont{Wang}},
  \bibinfo{author}{\bibfnamefont{R.~C.} \bibnamefont{Bialczak}},
  \bibinfo{author}{\bibfnamefont{E.}~\bibnamefont{Lucero}},
  \bibinfo{author}{\bibfnamefont{M.}~\bibnamefont{Mariantoni}},
  \bibinfo{author}{\bibfnamefont{M.}~\bibnamefont{Neeley}},
  \bibinfo{author}{\bibfnamefont{A.~D.} \bibnamefont{O'Connell}},
  \bibinfo{author}{\bibfnamefont{D.}~\bibnamefont{Sank}},
  \bibinfo{author}{\bibfnamefont{M.}~\bibnamefont{Weides}},
  \bibinfo{author}{\bibfnamefont{J.}~\bibnamefont{Wenner}},
  \bibnamefont{et~al.}, \bibinfo{journal}{Phys. Rev. B}
  \textbf{\bibinfo{volume}{84}}, \bibinfo{pages}{024501}
  (\bibinfo{year}{2011}).

\bibitem[{\citenamefont{Aumentado et~al.}(2004)\citenamefont{Aumentado, Keller,
  Martinis, and Devoret}}]{Aumentado}
\bibinfo{author}{\bibfnamefont{J.}~\bibnamefont{Aumentado}},
  \bibinfo{author}{\bibfnamefont{M.~W.} \bibnamefont{Keller}},
  \bibinfo{author}{\bibfnamefont{J.~M.} \bibnamefont{Martinis}},
  \bibnamefont{and} \bibinfo{author}{\bibfnamefont{M.~H.}
  \bibnamefont{Devoret}}, \bibinfo{journal}{Phys. Rev. Lett.}
  \textbf{\bibinfo{volume}{92}}, \bibinfo{pages}{066802}
  (\bibinfo{year}{2004}).

\bibitem[{\citenamefont{Schreier et~al.}(2008)\citenamefont{Schreier, Houck,
  Koch, Schuster, Johnson, Chow, Gambetta, Majer, Frunzio, Devoret
  et~al.}}]{Schreier}
\bibinfo{author}{\bibfnamefont{J.~A.} \bibnamefont{Schreier}},
  \bibinfo{author}{\bibfnamefont{A.~A.} \bibnamefont{Houck}},
  \bibinfo{author}{\bibfnamefont{J.}~\bibnamefont{Koch}},
  \bibinfo{author}{\bibfnamefont{D.~I.} \bibnamefont{Schuster}},
  \bibinfo{author}{\bibfnamefont{B.~R.} \bibnamefont{Johnson}},
  \bibinfo{author}{\bibfnamefont{J.~M.} \bibnamefont{Chow}},
  \bibinfo{author}{\bibfnamefont{J.~M.} \bibnamefont{Gambetta}},
  \bibinfo{author}{\bibfnamefont{J.}~\bibnamefont{Majer}},
  \bibinfo{author}{\bibfnamefont{L.}~\bibnamefont{Frunzio}},
  \bibinfo{author}{\bibfnamefont{M.~H.} \bibnamefont{Devoret}},
  \bibnamefont{et~al.}, \bibinfo{journal}{Phys. Rev. B}
  \textbf{\bibinfo{volume}{77}}, \bibinfo{pages}{180502}
  (\bibinfo{year}{2008}).

\bibitem[{\citenamefont{Friedman et~al.}(2000)\citenamefont{Friedman, Patel,
  Chen, Tolpygo, and Lukens}}]{Friedman}
\bibinfo{author}{\bibfnamefont{J.~R.} \bibnamefont{Friedman}},
  \bibinfo{author}{\bibfnamefont{V.}~\bibnamefont{Patel}},
  \bibinfo{author}{\bibfnamefont{W.}~\bibnamefont{Chen}},
  \bibinfo{author}{\bibfnamefont{S.~K.} \bibnamefont{Tolpygo}},
  \bibnamefont{and} \bibinfo{author}{\bibfnamefont{J.~E.}
  \bibnamefont{Lukens}}, \bibinfo{journal}{Nature}
  \textbf{\bibinfo{volume}{406}}, \bibinfo{pages}{43} (\bibinfo{year}{2000}).

\bibitem[{\citenamefont{van~der Wal et~al.}(2000)\citenamefont{van~der Wal, ter
  Haar, Wilhelm, Schouten, Harmans, Orlando, Lloyd, and Mooij}}]{van}
\bibinfo{author}{\bibfnamefont{C.~H.} \bibnamefont{van~der Wal}},
  \bibinfo{author}{\bibfnamefont{A.~C.~J.} \bibnamefont{ter Haar}},
  \bibinfo{author}{\bibfnamefont{F.~K.} \bibnamefont{Wilhelm}},
  \bibinfo{author}{\bibfnamefont{R.~N.} \bibnamefont{Schouten}},
  \bibinfo{author}{\bibfnamefont{C.~J. P.~M.} \bibnamefont{Harmans}},
  \bibinfo{author}{\bibfnamefont{T.~P.} \bibnamefont{Orlando}},
  \bibinfo{author}{\bibfnamefont{S.}~\bibnamefont{Lloyd}}, \bibnamefont{and}
  \bibinfo{author}{\bibfnamefont{J.~E.} \bibnamefont{Mooij}},
  \bibinfo{journal}{Science} \textbf{\bibinfo{volume}{290}},
  \bibinfo{pages}{773} (\bibinfo{year}{2000}).

\bibitem[{\citenamefont{Martinis et~al.}(2002)\citenamefont{Martinis, Nam,
  Aumentado, and Urbina}}]{Martinis2}
\bibinfo{author}{\bibfnamefont{J.~M.} \bibnamefont{Martinis}},
  \bibinfo{author}{\bibfnamefont{S.}~\bibnamefont{Nam}},
  \bibinfo{author}{\bibfnamefont{J.}~\bibnamefont{Aumentado}},
  \bibnamefont{and} \bibinfo{author}{\bibfnamefont{C.}~\bibnamefont{Urbina}},
  \bibinfo{journal}{Phys. Rev. Lett.} \textbf{\bibinfo{volume}{89}},
  \bibinfo{pages}{117901} (\bibinfo{year}{2002}).

\bibitem[{\citenamefont{Koch et~al.}(2007)\citenamefont{Koch, Yu, Gambetta,
  Houck, Schuster, Majer, Blais, Devoret, Girvin, and Schoelkopf}}]{Koch}
\bibinfo{author}{\bibfnamefont{J.}~\bibnamefont{Koch}},
  \bibinfo{author}{\bibfnamefont{T.~M.} \bibnamefont{Yu}},
  \bibinfo{author}{\bibfnamefont{J.~M.} \bibnamefont{Gambetta}},
  \bibinfo{author}{\bibfnamefont{A.~A.} \bibnamefont{Houck}},
  \bibinfo{author}{\bibfnamefont{D.~I.} \bibnamefont{Schuster}},
  \bibinfo{author}{\bibfnamefont{J.}~\bibnamefont{Majer}},
  \bibinfo{author}{\bibfnamefont{A.}~\bibnamefont{Blais}},
  \bibinfo{author}{\bibfnamefont{M.~H.} \bibnamefont{Devoret}},
  \bibinfo{author}{\bibfnamefont{S.~M.} \bibnamefont{Girvin}},
  \bibnamefont{and} \bibinfo{author}{\bibfnamefont{R.~J.}
  \bibnamefont{Schoelkopf}}, \bibinfo{journal}{Phys. Rev. A}
  \textbf{\bibinfo{volume}{76}}, \bibinfo{pages}{042319}
  (\bibinfo{year}{2007}).

\bibitem[{\citenamefont{Catelani
  et~al.}(2011{\natexlab{b}})\citenamefont{Catelani, Schoelkopf, Devoret, and
  Glazman}}]{Catelani2}
\bibinfo{author}{\bibfnamefont{G.}~\bibnamefont{Catelani}},
  \bibinfo{author}{\bibfnamefont{R.~J.} \bibnamefont{Schoelkopf}},
  \bibinfo{author}{\bibfnamefont{M.~H.} \bibnamefont{Devoret}},
  \bibnamefont{and} \bibinfo{author}{\bibfnamefont{L.~I.}
  \bibnamefont{Glazman}}, \bibinfo{journal}{Phys. Rev. B}
  \textbf{\bibinfo{volume}{84}}, \bibinfo{pages}{064517}
  (\bibinfo{year}{2011}{\natexlab{b}}).

\bibitem[{\citenamefont{Joyez et~al.}(1994)\citenamefont{Joyez, Lafarge,
  Filipe, Esteve, and Devoret}}]{Joyez}
\bibinfo{author}{\bibfnamefont{P.}~\bibnamefont{Joyez}},
  \bibinfo{author}{\bibfnamefont{P.}~\bibnamefont{Lafarge}},
  \bibinfo{author}{\bibfnamefont{A.}~\bibnamefont{Filipe}},
  \bibinfo{author}{\bibfnamefont{D.}~\bibnamefont{Esteve}}, \bibnamefont{and}
  \bibinfo{author}{\bibfnamefont{M.~H.} \bibnamefont{Devoret}},
  \bibinfo{journal}{Phys. Rev. Lett.} \textbf{\bibinfo{volume}{72}},
  \bibinfo{pages}{2458} (\bibinfo{year}{1994}).

\bibitem[{\citenamefont{Naaman and Aumentado}(2006)}]{Naaman}
\bibinfo{author}{\bibfnamefont{O.}~\bibnamefont{Naaman}} \bibnamefont{and}
  \bibinfo{author}{\bibfnamefont{J.}~\bibnamefont{Aumentado}},
  \bibinfo{journal}{Phys. Rev. B} \textbf{\bibinfo{volume}{73}},
  \bibinfo{pages}{172504} (\bibinfo{year}{2006}).

\bibitem[{\citenamefont{Court et~al.}(2008)\citenamefont{Court, Ferguson,
  Lutchyn, and Clark}}]{Court}
\bibinfo{author}{\bibfnamefont{N.~A.} \bibnamefont{Court}},
  \bibinfo{author}{\bibfnamefont{A.~J.} \bibnamefont{Ferguson}},
  \bibinfo{author}{\bibfnamefont{R.}~\bibnamefont{Lutchyn}}, \bibnamefont{and}
  \bibinfo{author}{\bibfnamefont{R.~G.} \bibnamefont{Clark}},
  \bibinfo{journal}{Phys. Rev. B} \textbf{\bibinfo{volume}{77}},
  \bibinfo{pages}{100501} (\bibinfo{year}{2008}).

\bibitem[{\citenamefont{Shaw et~al.}(2008)\citenamefont{Shaw, Lutchyn, Delsing,
  and Echternach}}]{Shaw}
\bibinfo{author}{\bibfnamefont{M.~D.} \bibnamefont{Shaw}},
  \bibinfo{author}{\bibfnamefont{R.~M.} \bibnamefont{Lutchyn}},
  \bibinfo{author}{\bibfnamefont{P.}~\bibnamefont{Delsing}}, \bibnamefont{and}
  \bibinfo{author}{\bibfnamefont{P.~M.} \bibnamefont{Echternach}},
  \bibinfo{journal}{Phys. Rev. B} \textbf{\bibinfo{volume}{78}},
  \bibinfo{pages}{024503} (\bibinfo{year}{2008}).

\bibitem[{\citenamefont{Wallraff et~al.}(2004)\citenamefont{Wallraff, Schuster,
  Blais, Frunzio, Huang, Majer, Kumar, Girvin, and Schoelkopf}}]{Wallraff}
\bibinfo{author}{\bibfnamefont{A.}~\bibnamefont{Wallraff}},
  \bibinfo{author}{\bibfnamefont{D.~I.} \bibnamefont{Schuster}},
  \bibinfo{author}{\bibfnamefont{A.}~\bibnamefont{Blais}},
  \bibinfo{author}{\bibfnamefont{L.}~\bibnamefont{Frunzio}},
  \bibinfo{author}{\bibfnamefont{R.-S.} \bibnamefont{Huang}},
  \bibinfo{author}{\bibfnamefont{J.}~\bibnamefont{Majer}},
  \bibinfo{author}{\bibfnamefont{S.}~\bibnamefont{Kumar}},
  \bibinfo{author}{\bibfnamefont{S.~M.} \bibnamefont{Girvin}},
  \bibnamefont{and} \bibinfo{author}{\bibfnamefont{R.~J.}
  \bibnamefont{Schoelkopf}}, \bibinfo{journal}{Nature}
  \textbf{\bibinfo{volume}{431}}, \bibinfo{pages}{162} (\bibinfo{year}{2004}).

\bibitem[{\citenamefont{Dolan}(1977)}]{Dolan}
\bibinfo{author}{\bibfnamefont{G.~J.} \bibnamefont{Dolan}},
  \bibinfo{journal}{Appl. Phys. Lett.} \textbf{\bibinfo{volume}{37}},
  \bibinfo{pages}{337} (\bibinfo{year}{1977}).

\bibitem[{\citenamefont{Houck et~al.}(2008)\citenamefont{Houck, Schreier,
  Johnson, Chow, Koch, Gambetta, Schuster, Frunzio, Devoret, Girvin
  et~al.}}]{Houck}
\bibinfo{author}{\bibfnamefont{A.~A.} \bibnamefont{Houck}},
  \bibinfo{author}{\bibfnamefont{J.~A.} \bibnamefont{Schreier}},
  \bibinfo{author}{\bibfnamefont{B.~R.} \bibnamefont{Johnson}},
  \bibinfo{author}{\bibfnamefont{J.~M.} \bibnamefont{Chow}},
  \bibinfo{author}{\bibfnamefont{J.}~\bibnamefont{Koch}},
  \bibinfo{author}{\bibfnamefont{J.~M.} \bibnamefont{Gambetta}},
  \bibinfo{author}{\bibfnamefont{D.~I.} \bibnamefont{Schuster}},
  \bibinfo{author}{\bibfnamefont{L.}~\bibnamefont{Frunzio}},
  \bibinfo{author}{\bibfnamefont{M.~H.} \bibnamefont{Devoret}},
  \bibinfo{author}{\bibfnamefont{S.~M.} \bibnamefont{Girvin}},
  \bibnamefont{et~al.}, \bibinfo{journal}{Phys. Rev. Lett.}
  \textbf{\bibinfo{volume}{101}}, \bibinfo{pages}{080502}
  (\bibinfo{year}{2008}).

\bibitem[{\citenamefont{Reed et~al.}(2010{\natexlab{a}})\citenamefont{Reed,
  Johnson, Houck, DiCarlo, Chow, Schuster, Frunzio, and Schoelkopf}}]{Reed1}
\bibinfo{author}{\bibfnamefont{M.~D.} \bibnamefont{Reed}},
  \bibinfo{author}{\bibfnamefont{B.~R.} \bibnamefont{Johnson}},
  \bibinfo{author}{\bibfnamefont{A.~A.} \bibnamefont{Houck}},
  \bibinfo{author}{\bibfnamefont{L.}~\bibnamefont{DiCarlo}},
  \bibinfo{author}{\bibfnamefont{J.~M.} \bibnamefont{Chow}},
  \bibinfo{author}{\bibfnamefont{D.~I.} \bibnamefont{Schuster}},
  \bibinfo{author}{\bibfnamefont{L.}~\bibnamefont{Frunzio}}, \bibnamefont{and}
  \bibinfo{author}{\bibfnamefont{R.~J.} \bibnamefont{Schoelkopf}},
  \bibinfo{journal}{Appl. Phys. Lett.} \textbf{\bibinfo{volume}{96}},
  \bibinfo{pages}{203110} (\bibinfo{year}{2010}{\natexlab{a}}).

\bibitem[{\citenamefont{Reed et~al.}(2010{\natexlab{b}})\citenamefont{Reed,
  DiCarlo, Johnson, Sun, Schuster, Frunzio, and Schoelkopf}}]{Reed2}
\bibinfo{author}{\bibfnamefont{M.~D.} \bibnamefont{Reed}},
  \bibinfo{author}{\bibfnamefont{L.}~\bibnamefont{DiCarlo}},
  \bibinfo{author}{\bibfnamefont{B.~R.} \bibnamefont{Johnson}},
  \bibinfo{author}{\bibfnamefont{L.}~\bibnamefont{Sun}},
  \bibinfo{author}{\bibfnamefont{D.~I.} \bibnamefont{Schuster}},
  \bibinfo{author}{\bibfnamefont{L.}~\bibnamefont{Frunzio}}, \bibnamefont{and}
  \bibinfo{author}{\bibfnamefont{R.~J.} \bibnamefont{Schoelkopf}},
  \bibinfo{journal}{Phys. Rev. Lett.} \textbf{\bibinfo{volume}{105}},
  \bibinfo{pages}{173601} (\bibinfo{year}{2010}{\natexlab{b}}).

\end{thebibliography}
\end{document}


\title{Supplemental Material for \\ Measurements of Quasiparticle Tunneling Dynamics in a Bandgap-Engineered Transmon Qubit}
\author{L.~Sun}
\affiliation{Department of Physics and Applied Physics, Yale University, New Haven, Connecticut 06520, USA}
\author{L.~DiCarlo}
\affiliation{Department of Physics and Applied Physics, Yale University, New Haven, Connecticut 06520, USA}
\affiliation{Kavli Institute of Nanoscience, Delft University of Technology, Delft, The Netherlands}
\author{M.~D.~Reed}
\author{G.~Catelani}
\affiliation{Department of Physics and Applied Physics, Yale University, New Haven, Connecticut 06520, USA}
\author{Lev~S.~Bishop}
\affiliation{Department of Physics and Applied Physics, Yale University, New Haven, Connecticut 06520, USA}
\affiliation{Joint Quantum Institute and Condensed Matter Theory Center, Department of Physics, University of Maryland, College Park, Maryland 20742, USA}
\author{D.~I.~Schuster}
\affiliation{Department of Physics and Applied Physics, Yale University, New Haven, Connecticut 06520, USA}
\affiliation{Department of Physics and James Franck Institute, University of Chicago, Chicago, Illinois 60637, USA}
\author{B.~R.~Johnson}
\affiliation{Department of Physics and Applied Physics, Yale University, New Haven, Connecticut 06520, USA}
\affiliation{Raytheon BBN Technologies, Cambridge, MA 02138, USA}
\author{Ge~A.~Yang}
\affiliation{Department of Physics and Applied Physics, Yale University, New Haven, Connecticut 06520, USA}
\affiliation{Department of Physics and James Franck Institute, University of Chicago, Chicago, Illinois 60637, USA}
\author{L.~Frunzio}
\author{L.~Glazman}
\author{M.~H.~Devoret}
\author{R.~J.~Schoelkopf}
\affiliation{Department of Physics and Applied Physics, Yale University, New Haven, Connecticut 06520, USA}
\maketitle

\subsection{Measured Devices}

All transmon qubits presented in the main text are measured by a coplanar waveguide cavity in a conventional circuit quantum electrodynamics architecture \cite{Wallraff} and the measurements are performed in a cryogen-free dilution refrigerator with a base temperature of about 20~mK. All devices are fabricated on a sapphire (Al$_2$O$_3$) substrate. Optical lithography followed by fluorine-based reactive ion etching is used to pattern the coplanar waveguide structure onto sputtered niobium. Electron-beam lithography is used to pattern the transmon qubits, which are deposited using a double-angle evaporation process. Table~\ref{Table:Devices} lists the parameters of the transmon qubits discussed in the main text. Qubits C, D, E, and F are on the same chip sharing a common readout resonator. All qubits have a quality factor $Q_f$ ranging between 50,000 and 90,000.

\begin{table}[b]
\begin{center}
\caption{Parameters of the transmon qubits. $\Delta$ is the superconducting gap of the electrodes of the Josephson tunnel junction, $f_{01}^{max}$ is the maximum qubit transition frequency between ground and the 1st excited state, $E_j^{max}$ is the maximum Josephson energy, $E_c$ is the charging energy, $g/2\pi$ is the coupling strength,  $f_c$ is the bare cavity frequency, $Q_c$ is the quality factor of the cavity, and $Q_f$ is the quality factor of the qubit.}
\begin{tabular}{c  c  c  c  c  c  c  c  c  c  c}
\hline
\hline
qubits&material&$\Delta$ ($\mu$eV)&quasiparticle trap&$\underset{(\mathrm{GHz})}{f_{01}^{max}}$&$\underset{(\mathrm{GHz})}{E_j^{max}}$
&$\underset{(\mathrm{MHz})}{E_c}$&$\underset{(\mathrm{MHz})}{g/2\pi}$& $\underset{(\mathrm{GHz})}{f_c}$
& $Q_c$ & $Q_f$\\
\hline
A [Fig.2(a)] & clean        & 180 &N & 8.250   & 36.4 & 248 & 145  & 6.810 & 1670 & 58,000\\
B [Fig.2(a)] & oxygen-doped & 280 &N & 5.158   & 12.4 & 300 & 190  & 6.912 & 6900 & 85,000\\
C [Fig.2(b)] & oxygen-doped & 280 &Y & 10.680  & 47.3 & 320 & 219  & 7.999 & 4100 & 65,000\\
D [Fig.2(b)] & oxygen-doped & 280 &Y & 10.112  & 42.5 & 320 & 237  & 7.999 & 4100 & 65,000\\
E [Fig.2(b)] & oxygen-doped & 280 &Y & 10.490  & 45.6 & 320 & 236  & 7.999 & 4100 & 65,000\\
F [Fig.2(b)] & oxygen-doped & 280 &Y & 10.560  & 46.2 & 320 & 215  & 7.999 & 4100 & 65,000\\
G [Figs.3-5]& oxygen-doped & 280 &Y & 10.558 & 30.6 & 500 & 280  & 8.072 & 500   & 57,000\\
\hline
\hline
\end{tabular}
\label{Table:Devices}
\end{center}
\end{table}

\subsection{Extraction of $\epsilon_0$, $\epsilon_1$, and $P_{odd}$ from Histograms}

Due to the switching of the qubit between the even and odd parities, the readout fidelity cannot be extracted directly from the measured histogram at one parity peak, but can easily be obtained by combining histograms at both parity peaks. For example, Fig.~S1 shows the cumulative probability after integrating the histograms taken at the even parity peak. $H_{even}$ and $L_{even}$ are the two cumulative probabilities below the readout threshold. Similarly, the counterparts of $H_{odd}$ and $L_{odd}$ at the odd peak are also measured. $H_{even}$, $L_{even}$, $H_{odd}$, and $L_{odd}$ are related to $\epsilon_0$, $\epsilon_1$, and $P_{odd}$ (readout false positives, false negatives, and the probability of the qubit at the odd parity, respectively) through the following equations:

\begin{equation}
L_{odd}=P_{odd}\epsilon_1+(1-P_{odd})(1-\epsilon_0),
\label{equ:H_odd}
\end{equation}
\begin{equation}
L_{even}=(1-P_{odd})\epsilon_1+P_{odd}(1-\epsilon_0),
\label{equ:H_even}
\end{equation}
\begin{equation}
H_{even}=H_{odd}=1-\epsilon_0.
\label{equ:L_even_odd}
\end{equation}

Solving these algebraic equations and using the measured values $H_{odd}=0.71$, $L_{odd}=0.47$, $H_{even}=0.71$, and $L_{even}=0.53$ gives $\epsilon_0\approx\epsilon_1=0.29$ and $P_{odd}=0.56$.

\begin{figure}
\includegraphics{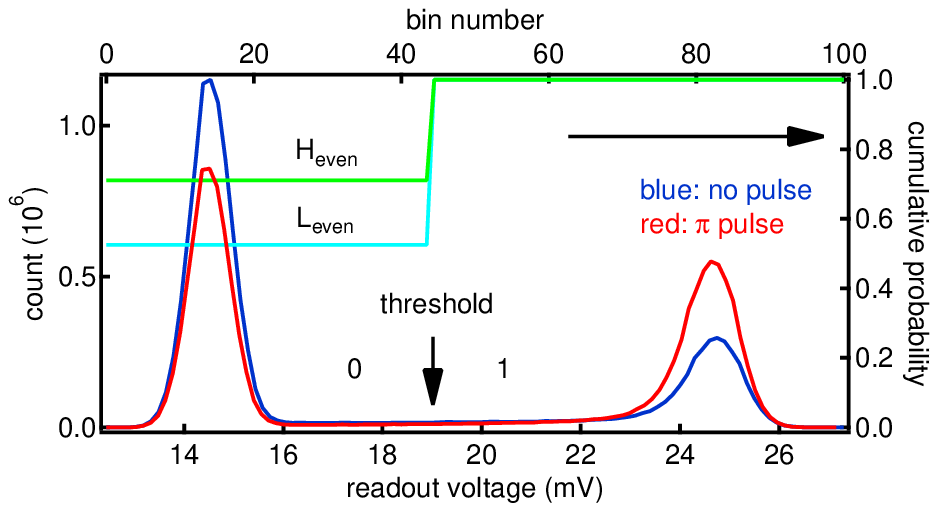}
\caption {Histogram of the readout at the even parity peak in Fig.~3(b) of the main text. A threshold $V_{th}=19$~mV is chosen to digitize the readout. Blue and cyan curves are the cumulative probability of the histograms after digitization. $H_{even}-L_{even}$, the cumulative probabilities difference below the threshold, gives the visibility. The cumulative probability of the histograms at the odd parity peak [Fig.~4(b) of the main text] are similar to those at the even parity peak here and are not shown.}
\label{fig:FigS1}
\end{figure}

\subsection{Spectrum of an RTS with Finite Readout Fidelity}

In our control experiment, a $\pi$ mask is used to randomly turn on and off the $\pi$ pulse applied to the qubit. This can imitate the success and failure of $\pi$ pulse on the qubit and allows us to test the sensitivity of our experiment to the qubit parity switching. For simplicity, here we again only consider the case when the $\pi$ pulse is applied at the odd parity frequency. If the time constants of the $\pi$ mask $\tau_{\pi 0}=\tau_{\pi 1}$ are large compared to qubit parity switching time [which is true even for the shortest $\tau_{\pi 0}=36~\mu$s in Fig.~5(a) based on the measured result $\tau_o\sim\tau_e<10~\mu$s from Fig.~5(b)], during $\tau_{\pi 0}$ ($\tau_{\pi 1}$) there are enough qubit parity switches to have statistical averages [Fig.~5(a) inset of the main text]:
\begin{equation}
A_\pi=P_{odd}(1-\epsilon_1)+(1-P_{odd})\epsilon_0,
\label{equ:A_pi}
\end{equation}
\begin{equation}
A_0=\epsilon_0.
\label{equ:A_pi}
\end{equation}
Therefore, the $\pi$ mask will generate an RTS with an amplitude
\begin{equation}
A_{con}=A_\pi-A_0=P_{odd}(1-\epsilon_1-\epsilon_0)=P_{odd}F<1.
\label{equ:A_con}
\end{equation}
Similarly, for the qubit parity switching rate $\Gamma_{oe}^{qp}$ measurement (Fig.~S2), the RTS amplitude is simply $A=A_1-A_0=(1-\epsilon_1-\epsilon_0)=F$ because of the absence of the $\pi$ mask. Note: the total variance of the readout signal in both control and qubit parity switching rate measurement is independent of $\tau_\pi$ and $\tau_{oe}$, but depends only on $\epsilon_0$, $\epsilon_1$, and $P_{odd}$. Each variance is thus constant. Explicitly, the total variance of the control experiment is $\sigma_{con}^2=[1-(A_\pi+A_0)/2](A_\pi+A_0)/2$ and the total variance of the parity switching rate experiment is $\sigma_{exp}^2=\{1-[P_{odd}A_1+(1-P_{odd})A_0]\}[P_{odd}A_1+(1-P_{odd})A_0]$. Therefore the total areas underneath different curves (the total spectral power) in Figs.~5(a) and 5(b) remain conserved, respectively.

The finite readout fidelity is only expected to generate a white noise background. Therefore, the theoretical predictions of the power spectral density (PSD) $S$ of an experiment with an RTS would be a sum of the spectra of an RTS and a white noise:
\begin{equation}
S=S_{RTS}+2t_s\sigma_{noise}^2,
\label{equ:spectrum_calculation}
\end{equation}
where $S_{RTS}$ is the PSD of the RTS, $t_s$ is the sampling time (10~$\mu$s), and $\sigma_{noise}^2$ is the total variance of the noise due to the finite readout fidelity. We have used the above equation to get the theoretical curves in Figs.~5(a) and 5(b) of the main text. Due to the finite measurement bandwidth, aliasing up to three times the Nyquist frequency in the control experiment and up to six times in the qubit parity switching rate measurement has been included in the theory curves. However, in the limit of a fast RTS signal, $\tau_o\sim\tau_e<t_s$, the RTS cannot be resolved any more and only a white noise is expected with an amplitude $S=2t_s\sigma_{exp}^2$ (again to maintain the total spectral power conserved) as indicated by the red dashed line in Fig.~5(b) of the main text. Note: the good agreement between the theoretical predictions and the control experiment also confirms the assumption that the parity switching time is small compared to the time constants of the $\pi$ mask and the parity switching only contributes to the white noise background because of the limited detection bandwidth.

\begin{figure}
\includegraphics{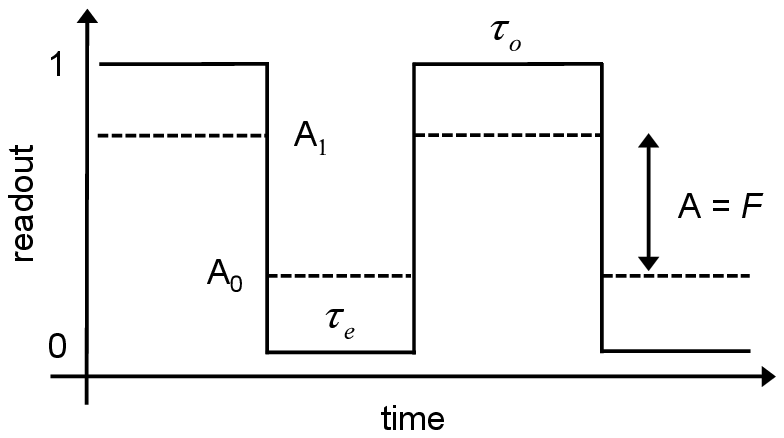}
\caption {Schematic of the qubit parity switching rate measurement. It is similar to the inset of Fig.~4(b) in the main text except that the RTS signal here is generated from the qubit parity switching. Due to the finite readout fidelity, the RTS amplitude is lowered to $A=F$.}
\label{fig:FigS_RTS}
\end{figure}

To get the instrumentation background noise and to ensure that it does not add correlation to the measurement, we perform readouts when applying neither the $\pi$ mask nor a $\pi$ pulse to the qubit and present its spectrum in Fig.~S3. In this case, because the qubit is always in the ground state and the readout is always performed at $f_{01}\approx 7$~GHz where the charge dispersion is negligible, the qubit parity switching does not make a difference in the readout and does not generate any RTS. Indeed, the spectrum is very close to the expected white noise background $S_{BG}=2t_s\epsilon_0(1-\epsilon_0)$ only due to the finite readout fidelity, except for small non-idealities like the $1/f$ noise.

\begin{figure}
\includegraphics{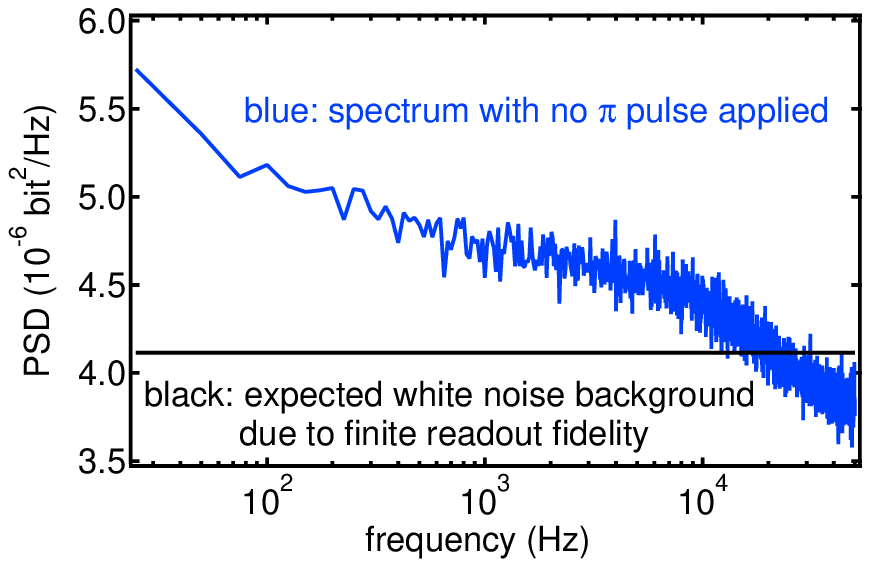}
\caption {Spectrum of readout with no $\pi$ pulse applied to the qubit. Blue: measured spectrum in linear scale with the qubit always in the ground state; black: expected white noise background due to finite readout fidelity. The measured spectrum is very close to the expected white noise background, demonstrating almost no correlation in the readout.}
\label{fig:FigS3}
\end{figure}